\documentstyle[amstex,12pt]{article}
\renewcommand{\paragraph}{}

\newcommand{\pr}{*_{\hbar}}
\newcommand{\tpr}{\, {\star}_{\hbar}\, }
\newcommand{\Mat}{{\mathsf Mat}}
\newcommand{\lon}{\longrightarrow}

\newcommand{\rar}{\rightarrow}

\newcommand{\End}{{\mathrm End}}

\newcommand{\Z}{{\Bbb Z}}
\newcommand{\p}{{\partial}}

\newcommand{\C}{{\Bbb C}}
\newcommand{\R}{{\Bbb R}}

\newcommand{\Id}{\mbox{Id}}
\newcommand{\Beq}{\begin{equation}}
\newcommand{\Eeq}{\end{equation}}
\newcommand{\Beqr}{\begin{eqnarray}}
\newcommand{\Eeqr}{\end{eqnarray}}
\newcommand{\Beqrn}{\begin{eqnarray*}}
\newcommand{\Eeqrn}{\end{eqnarray*}}
\newcommand{\Ba}{\begin{array}}
\newcommand{\Ea}{\end{array}}
\newcommand{\Bi}{\begin{itemize}}
\newcommand{\Ei}{\end{itemize}}
\newcommand{\Bc}{\begin{center}}
\newcommand{\Ec}{\end{center}}

\newcommand{\om}{\omega}

\newcommand{\sip}{\smallskip}
\newcommand{\bip}{\bigskip}

\begin{document}

\vspace{-3cm}

\title{The Moyal product is the matrix product} 
\author{ S.A.\ Merkulov\\
{\small Max Planck Institute for Mathematics in Bonn,}\vspace{-2mm}\\ 
{\small and
Department of Mathematics, University of Glasgow}}

\date{}
\maketitle

\begin{abstract}
This is a short comment on the Moyal formula for deformation quantization.
It is shown that the Moyal algebra of functions on the plane
is canonically isomorphic to  an algebra of matrices
of infinite size. 
\end{abstract}

\bip

\sip

{\bf 1. Deformation quantization.} Classical mechanical systems are 
mathematically described by symplectic manifolds, $(M,\om)$, and their
quantization is usually understood  as a functor
$$
  (M,\om) \lon {\mathsf a\ Hilbert\  space}\ H,
$$
which comes together with an association 
$$
\Ba{c}
{\mathsf a\ function\ }\ f\  
{\mathsf on}\  M \\ {\mathsf ``classical\ observable\mbox{''} }
\Ea 
\ \ \lon \ \ 
\Ba{c}
{\mathsf  an\  operator }\  \hat{f}\in
\End H[[\hbar]]\\ {\mathsf ``quantum\ observable \mbox{''}}
\Ea
$$
satisfying the conditions 
\Bi
\item[(i)] $\widehat{1}=\Id$,\\
\item[(ii)] $\widehat{\{f,g\}}=\lim_{\hbar \rar 0}\frac{i}{\hbar}
\left({\widehat{f}\, \widehat{g} - \widehat{g}\, \widehat{f}}
\right)$,\\
\item[(iii)] ${\mathsf complex \ conjugation} \stackrel{\widehat{\ \ } }
{\lon}
{\mathsf transition\  to\  the\  adjoint}$,
\Ei
where $\{\ , \ \}$ is the Poisson bracket, 
$$
\{f,g\}:= \om^{-1}(df,dg).
$$
Most symplectic manifolds arising in classical mechanics are total
spaces of the cotangent bundles,
$(T^*P, {\mathsf standard\ symplectic\ form})$, to some $n$-dimensional
{\em configuration}\, manifolds $P$, and the above association is often plagued
with the ordering choice ambiguity.

\sip

Deformation quantization \cite{BF} offers a very different scheme
 in which the classical
data, $(C^{\infty}(M),\om)$, is mapped upon quantization to itself as a
set but {\em not}\, as a ring. The usual
commutative product of functions on $M$ gets replaced by
 a new {\em deformed}\,  product, 
$$
f*_{\hbar} g= fg +\sum_{n=1}^{\infty} \hbar^n D^n(f,g), \ \ \ f,g\in 
C^{\infty}(M)[[\hbar]],
$$
which is supposed to satisfy the conditions,
\Bi
\item[(i)] $D_n$ are $\R[[\hbar]]$-linear bidifferential operators of 
finite total order,\\
\item[(ii)] $1*_{\hbar} f=f*_{\hbar}1=f$,\\
\item[(iii)] ${\{f,g\}}=\lim_{\hbar \rar 0}\frac{1}{\hbar}
\left({f}*_{\hbar} {g} - {g}*_{\hbar}{f}\right)$, 
\item[(iv)] $(f\pr g)\pr h=f\pr (g\pr h)$,
\Ei
for all $f$, $g$ and $h$ in $ C^{\infty}(M)[[\hbar]]$.  The deformed
product is associative, but no more commutative. The above mentioned
ordering choice ambiguity shows itself again, this time in
the form of an equivalence relation among star products: 
$\pr \sim \tpr$ if there exist
$\R[[\hbar]]$-linear differential operators $Q_n: C^{\infty}(M)\rar
C^{\infty}(M)$ such that
$$
Q(f\tpr g) = (Qf)\pr (Qg),  \ \ \ \  \forall f,g\in C^{\infty}(M)[[\hbar]],
$$
where $Q=\Id + \sum_{n\geq 1} Q_n$. 

\sip

The spectral theory of observables can be studied, within the deformation
quantization framework, via the so-called star-exponentials \cite{BF}.

\bip

\sip

{\bf 2. The Moyal product.} This is the simplest example of deformation
quantization. The symplectic input is $\R^{2n}$ with its 
standard 2-form $\om= \sum_{i=1}^n dx^i\wedge dp_i$ and the deformed
product is given by the following formula\footnote{Strictly speaking,
we have to replace $\hbar$  by ${i\hbar}$ everywhere in this text
to make things consistent with the usual physics conventions.}, 
$$
f\pr g := \left. e^{\sum_{a=1}^n \frac{\hbar}{2}\left(
\frac{\p^2}{\p x^a\p \tilde{p}_a} -
\frac{\p^2}{\p p_a\p
\tilde{x}^a}\right)}
f(x^b,p_b) g(\tilde{x}^c,\tilde{p}_c)\right|_{{x^a=\tilde{x}^a}\atop
p_a=\tilde{p}_a}.
$$
On the plane, $\R^2$,  the Moyal product simplifies,
$$
f\pr g= \sum_{n=0}^{\infty} \frac{\hbar^n}{2^n n!} \sum_{k=0}^n
\binom{n}{k} (-1)^k \frac{\p^n f}{\p x^{n-k}\p p^k} \frac{\p^n
g}{\p x^k \p p^{n-k}}.
$$ 
Quantum mechanically, this product corresponds to the Weyl  (symmetric)
ordering in the  ring of observables $C^{\infty}(\R^2)[[\hbar]]$. 
The equivalent product,
\Beqrn
f \tpr g &:=& e^{\frac{\hbar}{2} \frac{\p^2}{\p x \p p}}\left(
(e^{-\frac{\hbar}{2} \frac{\p^2}{\p x \p p}} f)\pr
(e^{-\frac{\hbar}{2} \frac{\p^2}{\p x \p p}} g)\right)\\
&=&  \sum_{n=0}^{\infty} \frac{\hbar^n}{n!} \frac{\p^n f}{\p x^n}
\frac{\p^n g}{\p p^n},
\Eeqrn
corresponds to the standard (non-symmetric) ordering.

\sip

From now on we assume that the symbols $\pr$ and $\tpr$ stand for
 the products just defined. As the space of
observables we take the formal ring, $k[[x,p,\hbar]]$,
where $k$ is a field containing rational numbers (for example $\R$ or $\C$).

\bip

\sip

{\bf 3. Matrix algebra.} Let $\Z^{\geq 0}$ be the ring of non-negative
integers, $R$ the  algebra of formal power series  $k[[\hbar]]$ 
and $I$ the maximal
ideal in $R$.   Denote by ${\Mat}_{\infty}^{\hbar}$ the algebra of all
matrices, $(a_{ij})_{i,j\in \Z^{\geq 0}}$, such that
$$
a_{ij}\in \left\{\Ba{cl} R, & \mbox{if}\ i\geq j,\\
I^{j-i}, & \mbox{if}\ i< j. \Ea \right.
$$

Pictorially, 
$$
{\Mat}_{\infty}^{\hbar} = \left( \begin{matrix}
R & I & I^2 & I^3 & I^4 & \ldots \\
R & R & I   & I^2 & I^3 & \ldots \\
R & R & R   & I   & I^2 & \ldots \\
R & R & R   & R   & I & \ldots \\
R & R & R   & R   & R & \ldots \\
\ldots & \ldots & \ldots   & \ldots   & \ldots & \ldots \\
\end{matrix}
   \right)
$$
This display makes it obvious that the usual matrix multiplication in 
${\Mat}_{\infty}^{\hbar}$ is  well-defined.

\sip

The algebra ${\Mat}_{\infty}^{\hbar}$ is freely generated as an $R$-module
by the matrices, $E_{a,b}$, $a,b\in \Z^{\geq 0}$, whose $ij$-th entry
is, by definition, given by
$$
(E_{a,b})_{ij} =\left\{ \Ba{cl} \frac{(b+k)!}{k!}\hbar^{b}, & 
\mbox{if}\ i=a+k, j=b+k, k=0,1,2,\ldots, \\
0, & \mbox{otherwise}.\Ea\right.
$$
One may check that their matrix product is given by
\begin{equation}\label{E}
E_{a,b}E_{c,d}= \sum_{n=0}^{\min(b,c)} \frac{\hbar^nb!c!}{n!(b-n)!(c-n)!}
E_{a+c-n,b+d-n}.
\end{equation}

\sip

\bip

{\bf 4. Proposition.} {\em There is a canonical isomorphism of 
associative algebras,
$$
\phi: \left(k[[x,p,\hbar]], \tpr \right) \lon {\Mat}_{\infty}^{\hbar}
$$
given on the generators as follows,
$$
\phi(p^ax^b)= E_{a,b}.
$$
}

\bip

This result implies in turn the main claim of this paper.

\bip

{\bf 5. Main Theorem.} {\em The Moyal algebra 
$\left(k[[x,p,\hbar]], \pr \right)$ is canonically isomorphic to the 
matrix algebra ${\Mat}_{\infty}^{\hbar}$. On the generators,
the isomorphism $\psi$ is given by
$$
\psi(p^ax^b)= \sum_{n=0}^{\min(a,b)} \frac{\hbar^na!b!}{2^nn!(a-n)!(b-n)!}
E_{a-n,b-n}.
$$
}

\sip

\bip

{\bf 6. Example.} To see how it all works, let us consider
two monomials, $p$ and $x^2$, with
$$
p\pr x^2 = px^2 - \hbar x.
$$
We have
$$
\psi(p)= E_{1,0}= \left( \begin{matrix}
0 & 0 & 0 & 0 & 0 & \ldots \\
1 & 0 & 0 & 0 & 0 & \ldots \\
0 & 1 & 0 & 0   & 0 & \ldots \\
0 & 0 & 1 & 0   & 0 & \ldots \\
0 & 0 & 0 & 1   & 0 & \ldots \\
\ldots & \ldots & \ldots   & \ldots   & \ldots & \ldots \\
\end{matrix}
   \right)
$$
and
$$
\psi(x^2) = E_{0,2} = \left( \begin{matrix}
0 & 0 & \frac{2!}{0!}\hbar^2 & 0 & 0 & \ldots \\
0 & 0 & 0 & \frac{3!}{1!}\hbar^2 & 0 & \ldots \\
0 & 0 & 0 & 0   & \frac{4!}{2!}\hbar^2 & \ldots \\
0 & 0 & 0 & 0   & 0 & \ldots \\
0 & 0 & 0 & 0   & 0 & \ldots \\
\ldots & \ldots & \ldots   & \ldots   & \ldots & \ldots \\
\end{matrix}
   \right)
$$
implying
\Beqrn
\psi(p)\psi(x^2) & = & \left( \begin{matrix}
0 & 0 &     0                & 0 & 0 & \ldots \\
0 & 0 & \frac{2!}{0!}\hbar^2 & 0 & 0 & \ldots \\
0 & 0 & 0 & \frac{3!}{1!}\hbar^2 & 0 & \ldots \\
0 & 0 & 0 & 0   & \frac{4!}{2!}\hbar^2 & \ldots \\
0 & 0 & 0 & 0   & 0 & \ldots \\
\ldots & \ldots & \ldots   & \ldots   & \ldots & \ldots \\
\end{matrix}
   \right) \\
&=& E_{1,2} \\
&=& (E_{1,2} + \hbar E_{0,1}) - \hbar E_{0,1}\\
&=& \psi(px^2) - \hbar \phi(x)\\
&=& \psi(p\pr x^2).
\Eeqrn

\sip

Analogously,
\Beqrn
\psi(x^2)\phi(p) & = & \left( \begin{matrix}
 0 & \frac{2!}{0!}\hbar^2 & 0 & 0 & 0 &\ldots \\
 0 & 0 & \frac{3!}{1!}\hbar^2 & 0 & 0 &\ldots \\
 0 & 0 & 0   & \frac{4!}{2!}\hbar^2 & 0 & \ldots \\
 0 & 0 & 0   & 0 & \frac{5!}{3!}\hbar^2 &  \ldots \\
 0 & 0 & 0 & 0   & 0 & \ldots \\
\ldots & \ldots & \ldots   & \ldots   & \ldots & \ldots \\
\end{matrix}
   \right) \\
&=& E_{1,2} + 2\hbar E_{0,1} \\
&=& (E_{1,2} + \hbar E_{0,1}) + \hbar E_{0,1}\\
&=& \psi(px^2) + \hbar \phi(x)\\
&=& \psi( x^2 \pr p),
\Eeqrn
which completes the check.

\bip

{\bf 7. Proof of the Proposition.} Once the matrices
$E_{a,b}$ with the properties (\ref{E}) are explicitly written down,
the proof becomes obvious. Thus we have only to motivate our
definition of $(E_{a,b})$s and, probably, indicate how one might check the 
key properties (\ref{E}) without serious calculations:
\Bi
\item[1)] In the first place, we have read these matrices  out of the quantum
space of the $n$-tuple point, $x^n=0$, \cite{Me}; the existence of
the isomorphisms $\phi$ and $\psi$ can be deduced from the 
projective limit of that construction. 
\item[2)] Alternatively, one may check that the $\tpr$-product of the
 functions
$$
g_{a,b}:= \frac{p^a x^b}{b!\hbar ^b}e^{-\frac{px}{\hbar}}, \ \ a,b=0,1,2,\dots,
$$
is well-defined and is given by
$$
g_{a,b}\tpr g_{c,d} = \delta_{bc} g_{a,d},
$$
where
$$
\delta_{bc}=\left\{\Ba{cl} 0, & \mbox{if}\ b\neq c,\\
1, & \mbox{if}\ b=c. \Ea \right.
$$
This fact is very easily established from the obvious differential equations
satisfied by $g_{a,d}$ and $g_{a,b}\tpr g_{c,d}$ as well as their
initial values (modulo the factor $p^ax^d$) at $x=p=0$. 

\sip

Next one notices that
$$
p^a x^b = \sum_{n=0}^{\infty} \hbar^b \frac{(b+n)!}{n!}g_{a+n,b+n},
$$
explaining both the isomorphism $\phi$ in the Proposition, and the
properties (\ref{E}) of the matrices $E_{a,b}$.
\Ei
\hfill $\Box$

\sip

\bip

{\bf 8. Proof of the Main Theorem.} One may deduce this statement
directly from the Proposition. Indirectly, one notices that the functions,
\Beqrn
h_{a,b} &=& e^{-\frac{\hbar}{2}\frac{\p^2}{\p x \p p}} \left(g_{a,b}\right) \\
&=& \frac{2}{b!\hbar^b} e^{-\frac{2xp}{\hbar}}
e^{-\frac{\hbar}{4}\frac{\p^2}{\p x \p p}} \left(p^a x^b\right) 
\Eeqrn
satisfy
$$
h_{a,b}\pr h_{c,d} = \delta_{bc} h_{a,d},
$$
and that one has
$$
e^{-\frac{\hbar}{2}\frac{\p^2}{\p x \p p}}\left( p^a x^b\right) =
\sum_{n=0}^{\infty} \hbar^b \frac{(b+n)!}{n!}h_{a+n,b+n}.
 $$
This explains the structure of the isomorphism $\psi$.
\hfill $\Box$



{\small
\begin{thebibliography}{99}


\bibitem[BFFLS]{BF} Bayen, F,  Flato, M,  Fronsdal, C.,
Lichnerovich, A. and  Sternheimer, D.: { Deformation quantization,
I \& II}, Ann.\ Phys.\ {\bf 111}, 61-110 \& 111-151 (1978)


\bibitem[Mo]{Mo}  Moyal, J.E.: { Quantum mechanics as a 
statistical theory}, Proc.\ Cambridge Phil.\ Soc.\ {\bf 45}, 
99-124 (1949)


\bibitem[Me]{Me} Merkulov, S.A.: {Deformation quantization of the $n$-tuple
point}, Commun.\ Math.\ Phys. {\bf 205}, 369-375 (1999)

\end{thebibliography}
}
\end{document}